# Big Data and Large Numbers: Interpreting Zipf's Law

Horia-Nicolai Teodorescu
Technical University of Iasi
Iasi, Romania
and Institute of Computer Science of the Romanian Academy
Iasi / Bucharest, Romania

*Abstract*—It turns out that some empirical facts in Big Data are the effects of properties of large numbers. Zipf's law "noise" is an example of such an artefact. We expose several properties of the power law distributions and of similar distribution that occur when the population is finite and the rank and counts of elements in the population are natural numbers. We are particularly concerned with the low-rank end of the graph of the law, the potential of noise in the law, and with the approximation of the number of types of objects at various ranks. Approximations instead of exact solutions are the center of attention. Consequences in the interpretation of Zipf's law are discussed.

## I. Preliminaries

The large number of models for statistical processes found in various domains such as linguistics, economy, demographics, computer science (file sizes, download times), and networks (Internet topology, Web graph) is disconcerting (Mitzenmacher). Mitzenmacher lists "…Pareto: income distribution, stock prices; Zipf-Auerbach: city sizes; Zipf-Estouf: word frequency; Lotka: bibliometrics; Yule: species and genera; Mandelbrot: economics/information theory". The corresponding distributions are as follows: Pareto distribution $p(X \geq x) = (x/k)^{-\alpha}$ ; power law distribution $p(X \geq x) \sim cx^{-\alpha}$ ; log-complementary cumulative distribution function $\ln p(X \geq x) = \alpha(\ln k - \ln x)$ (Mitzenmacher). We are concerned with some properties of these distributions for the case of discrete variables and with applications primarily to texts. While we focus on the power laws, the discussion remains largely valid for whatever variant of process that has a probability distribution function defined for discrete variables and tending to zero at infinity, including the log-normal distribution and Mandelbrot distribution.

In this paper we are interested in rank distributions where ranks are natural numbers, $r = 1, 2, \ldots$ The population is assumed whatever large but finite; the population count will be denoted by $N$. The population is a multiset, possibly with several elements of the population of the same type. The types are discrete and the number of types is lower than the number of elements in the population. Throughout the paper the population or subpopulations will also be named collections. An example of population and subpopulations of interest is given in (Teodorescu M., 2017), (Teodorescu M., 2017, 2018), who analyzed Zipf's law usefulness in patent sub-corpora comparisons for "the US-granted patent abstracts corpus, the US-granted patent titles corpus, and the US-granted patent claims corpus (all patents issued between 2007 and November 2017)". Further, the population may be a set of towns and cities and the types are the counts of their populations, or the population is composed by the people of a country and the types are the ranges of their income, or the population is composed by the inventors who patented in a country and the types are the counts of their patents. Further examples include tourist flows (Blackwell et al.), (Lin and Lee), firms that go bankrupt (Fujiwara 2003), software component sizes (Sharma and Pendharkar, 2022), patents as indicators of innovation (O'Neale 2012), (Teodorescu M., 2018), statistics of the participation in Internet communities (Tenorio-Fornés et al.), or sentences in a text. We are concerned in this article with populations of words (or lemmas) from specified texts; the types are the words. For introductory details see also (Corral et al. 2015); for a good example of a comparison of Zipf's laws occurring in various types of texts (Brown Corpus the patent titles corpus, the patent abstracts corpus, and the patent claims corpus), see (Teodorescu M., 2018), Figure 3, at page 49 in that article.

We count the number of elements of type $t$ in the population, $n(t)$, and sort these counts descendingly; the type with the largest count is ranked first, and the count will be denoted $n(r = 1)$ or $n(1)$. We admit that the studied population may be a sub-multiset of a larger multiset $U$ and that some types in $U$ may lack in the studied population; however, we are interested in ranks up to $n(r) \geq 1$, i.e., ranks of the elements of population, not of $U$.

There is some confusion in the literature between two forms of Zipf's law. The first is produced by randomly generating a finite number $N$ of real numbers with a distribution $\sim \frac{1}{r^a}$, counting how many of them, $n(b_j)$, are in a set of bins $b_j$, starting at some value $x_j$, and then representing $n(b_j(x_j))$ as a function $n(x_j)$. Especially when $N$ is relatively small compared with $\frac{1}{x_{j_m}}$, where $x_{j_m}$ is the largest $x_j$, the number of elements counted in the bins with large $x_j$ has a large chance to depart from the number expected for $N \to \infty$, hence the "noisy" end in most papers using with this form of generating Zipf's law. An example is (Newman, 2005). In that paper, the author uses $10^6$ randomly generated values produced with a power law, then places them into bins of range 0.1, and comments the obtained graph as "Notice how noisy the results get in the tail towards the right-hand side of the panel. This happens because the number of samples in the bins becomes small and statistical fluctuations are therefore large as a fraction of sample number." But by using bins, one creates "types" of elements, one type per bin. The noise occurs because the abscissa is on values, not on ranks; the obtained graph is count-per-value, not count-per-rank. Yet, for $N \to \infty$, the counts per value tend to the count per rank. When large noise occurs in this form of Zipf's law, it is just an indicator that the number or ranks considered is too large for the number of values in the population.

The second form of Zipf's law is obtained as a count per rank law; it is the natural representation when types are enforced by the problem; in texts, words or phrases are natural types. Their counts versus the corresponding ranks are found to obey Zipf's law. It is this this form of Zipf's law we address; this form is natural when the elements in the population have qualitative, non-numerical features. This form of the law also seems to have a noisy end, as discussed in the next section.

In language and graph statistics one is interested in counts that are natural numbers and ranks are natural numbers. Therefore, the power laws have the form

$$n(r) = \frac{A}{r^a}, a \in \mathbf{R}, a > 0, A \in \mathbf{R}, \ A > 1. \quad (1)$$

Because the counts $n(r)$ are integers, the equality in (1) is not generally possible; therefore, one should interpret (1) as the rounded $n(r) = \left[\frac{A}{r^a}\right]$, or the transformation of the right side to an integer can be done using the floor or the ceiling functions. Another condition derived from the concept of rank and from the rank ordering is that, for any $h < j$, $n(h) \geq n(j)$. This imposes that $a \geq 0$, where the case $a = 0$ is of no interest (all types have the same rank).

The definition (1) is reminiscent of Riemann function $\zeta(s) = \sum_{r=1}^{\infty} \frac{1}{r^s}$, $s \in \mathbf{C}$. We will use several elementary properties of this function, among others the values of $\zeta(2)$ and $\zeta(3)$. Because the populations we are concerned with are finite, the sums of counts for all ranks are partial sums of $\zeta$ up to the maximal rank $r_M$,

$$S_{r_M}(a) = \sum_{r=1}^{r_M} n(r) = A \sum_{r=1}^{r_M} \frac{1}{r^a} \quad (2)$$

and are finite; we are not concerned with the convergence of $S_{r_M}$, although we use the asymptotic approximation according to $\zeta(s)$. For larger even values of the exponent, the partial series $S_{r_M}$ tend to the values computed as in (Ciaurri et al., 2015). Notice that $S_{r_M}(a) = A \cdot H_{r_M}^{(a)}$, where $H_m^{(a)}$ denotes the harmonic number with exponent $a$; standard harmonic numbers are $H_m^{(1)}$. In the examples, we will use some values computed with Mathematica®, for example for $a = 3/2$, $\zeta(3/2) = 2.61237 \ldots$ (Wolfram).

This article is based on a set of assumptions that may be disputable. The most important is that the type of the objects is well determined (not a random variable). This may be untrue. For example, if the types are given by a set of ranges of how much people spend in a given year, probably most people do not know precisely the figures, thus the types become blurred. Other applications with imprecise ranking pertain to preference theory and psychology (Thurstone), (Li et al., 2022), (Barrientos et al.). As a consequence, in statistics, the counts of types are estimations, not numbers; they should be better described by stochastic variables. This in turn makes the ranks imprecise. Studying rank distributions with imprecise ranks is an important domain of investigation in communications and other fields, see for example (Li et al. 2009). Nevertheless, in this article the results are based on natural number counts and ranks. We also assume that the distribution is a power law. However, we will allow noise in the power law; moreover, we briefly deal with a rank-dependent exponent in the form $1/r^{(a+\mu/r)}$. Other similar laws, such as the log-normal, produce similar results to the empirical ones obtained in various domains, departing from Zipf's law.

## II. PROPERTIES INDUCED BY THE INTEGER COUNTS AND RANKS

Essential differences occur between the real valued power law of real variable and the case when the variable and the ranks are natural numbers. In the last case, the natural numbers impose restrictions and induce specific properties. One of the consequences is that the "noisy end" of the graph of power law populations is not true noise.

Consider the basic model of power law (1) for a population of $N$ elements, each count corresponding to a different type. The total number of elements is

$$N = \sum_{r=1}^{r_M} \frac{A}{r^a} = A(\zeta(a) - \psi(a, r_M)), \quad (3)$$

where $A$ is a constant, $A = N/S_{r_M} = N/\sum_{r=1}^{r_M} \frac{1}{r^a}$ and $\psi(a, r_M)$ will be named the zeta-error function

$$\psi(a, r_M) = \zeta(a) - S_{r_M}(a) \quad (4)$$

and is used for ease of explanation. The model is however incompatible with the conditions $n(r) \in \mathbf{N}, r \in \mathbf{N}, a > 0, a \in \mathbf{R}$, because the partial sums are not natural numbers (not even for $a \in \mathbf{N}$). Consequently, the actual counts may be either modeled as rounded numbers, as discussed in (Teodorescu HN, 2023) for $a = 1$, or as variables with attached probabilities. For example, the count for rank $r$ may have two values, $\left\lfloor\frac{A}{r^a}\right\rfloor$ with probability $p_1$, or $\left\lceil\frac{A}{r^a}\right\rceil$ with probability $p_2$. For applications, one may choose $p_1 = p_2 = 0.5$, or one may chose $p_1 = \int_r^{r+0.5} \frac{B}{r^a} dr$, $p_2 = \int_{r+0.5}^{r+1} \frac{B}{r^a} dr$, $B = 1/\int_r^{r+1} \frac{1}{r^a} dr$. For now, we will use the adjusted model based on rounding,

$$n(r) = \left[\frac{A}{r^a}\right]. \quad (5)$$

The next problem is that that for large enough values of $r$, one obtains

$$n(r) = \left[\frac{A}{r^a}\right] = \left[\frac{A}{(r+1)^a}\right] = n(r+1). \quad (6)$$

But this equality contradicts the unicity of ranks in statistics where only "entire" elements are counted, such as entire words, not fractions of them. Therefore, two or several successive ranks have to be allowed to correspond to the same count. It is why in the Preliminaries section we allowed for the condition that for any two ranks $h < j$, $n(h) \geq n(j)$ instead of using the strict condition $n(h) > n(j)$.

We may be interested if there is a threshold rank, $r_0$, such that for ranks $r < r_0$ the model predicts a single rank for a given count, while for ranks $r > r_0$ the model predicts the

possibility of two or more ranks with the same count. Equivalently, the conditions defining $r_0$ are

$$r < r_0 \rightarrow \frac{A}{r^a} \geq \frac{A}{(r+1)^a} + 1, \quad (7)$$

$$\forall r \geq r_0, j \geq 1: \frac{A}{(r+j)^a} - \frac{A}{(r+j+1)^a} < 1. \quad (8)$$

The first condition, (7), leads to the equation $(A, r \in \mathbb{N})$

$$A(r+1)^a \geq r^a(A + (r+1)^a), \quad (9)$$

with $r_{01}$ the smallest value satisfying the above. The minimal value of $r$ in the second condition (8) leads to $r_{02}$, if $r_{02} < r_M$. However, it is possible that the two conditions produce different values, $r_{01} \neq r_{02}$; we chose $r_0 = r_{01}$. Both conditions have to be satisfied for $r_0$ has the sense of the maximal range with a count guaranteed different from the next one.

To simplify the discussion, for the count at rank $r$ we use in the remaining part of the paper the definition based on the floor function,

$$n(r) = \left\lfloor \frac{A}{r^a} \right\rfloor. \quad (10)$$

Figure 1 shows a surrogate (simulated) Zipf's-like law obtained for $M = 10^6, a = -2$; in red is the law without considering the rounding of the number of elements.

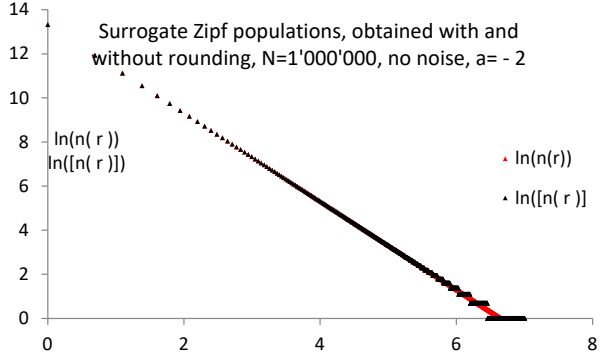

Fig. 1. Surrogate Zipf's law.

Fig. 2 shows the variation of the number of ranks with the same number of elements (due to rounding), for two values of $M$.

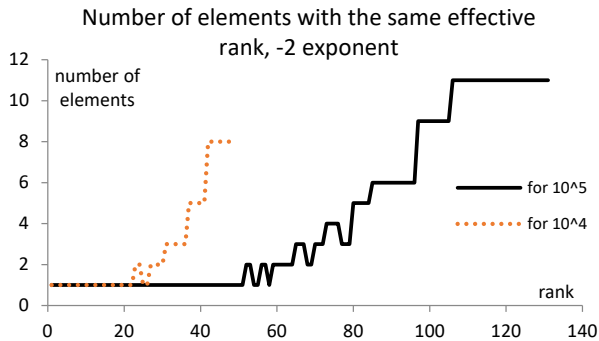

Fig. 2. Number of ranks with the same count.

III. MERGING TWO POPULATIONS WITH POWER LAW RANKS

Let be two populations $C_1, C_2$ of discrete values from the same set (e.g., words from the same language). Let $m_j = m(r_j)$ and $n_h = n(r_h)$ be the number of elements of rank $r_j$ and $r_h$ in the two populations. Denote by $w$ an element of one of the populations. Then, it may exist $w \neq w'$ such that $m(w) = m(w')$. In this case, $w, w'$ will be assigned successive ranks (either one being the first); similarly for whatever number of elements with the same number. For example, if "yes" and "no" both occur 2 times, and "maybe" it the only word that occurs 3 times and has rank 8, then the ranks assigned may be $r(yes) = 4, r(no) = 5$.

Denote by $r_{jM}$ and $r_{hM}$ the largest ranks satisfying $m_{jM} = 1$ and $n_{hM} = 1$. Then:

If $m_{jM} = m_{hM}$, the number of elements in $C_1 \cup C_2$ (where $C_1, C_2$ and $C_1 \cup C_2$ are multisets and $\cup$ denotes merging of the multisets) with only one occurrence in $C_1 \cup C_2$ is at most the sum of the numbers of elements with a single occurrence in the initial sets $C_1, C_2$. In linguistic terms, the number of hapax legomena in the merged multiset is at most the sum of hapaxes in the initial collections $C_1, C_2$. This is because there are no new words (others than those in $C_1, C_2$) in $C_1 \cup C_2$, while some of the hapaxes may occur in both collections and thus be no more hapaxes in the merged collection.

Assume (unrealistically) that $C_1$ and $C_2$ include precisely the same elements and respect the conditions:

$$n_{11} > n_{12} > n_{21} > n_{22} > \cdots > n_{1h} > n_{2h} > n_{1h+1} = n_{1h+2} = n_{2h+1} = n_{2h+2} > n_{1h+3} = \cdots = n_{2h+3} = \cdots = n_{2h+u} > n_{1h+u} = \cdots \quad (11)$$

In words, in both sets, there are a number of elements up to $h$ having for any rank $1 \ldots h$ a single word with the same count. In addition, the counts are such that the counts for elements in $C_2$ are between the successive counts in $C_1$. For ranks higher than $h$, there are at least two words with the same counts in each collection, meaning that in the merged collection there will be at least four distinct elements with the same count, thus having successive ranks. Because we assumed that the elements of $C_1$ and $C_2$ are distinct, we will have $2h$ elements in $C_1 \cup C_2$ having no competitor to the same rank (no element with the same count, because we assumed that $n_{12} > n_{21} > n_{22}$ etc). Supposing that while different $n_{1h} > n_{2h}$, $n_{1h} \approx n_{2h}$, $\ln r_{2 \times h}(C_1 \cup C_2) - \ln r_1(C_1 \cup C_2) = \ln 2h - \ln 1 = \ln 2 + \ln h$ and $\ln n_{11} - \ln n_{2 \times h}(C_1 \cup C_2) \approx \ln n_{11} - \ln n_{1h}(C_1)$. We explicitly used the notation $(C_1 \cup C_2)$ to show that the rank or count is for the merged set. Consequently, the slope of the line segment $y = \ln(count) = a + b \ln(rank)$ changes from $\frac{\ln n_{11} - \ln n_{1h}}{\ln h}$ to $\frac{\ln n_{11} - \ln n_{1h}}{\ln 2 + \ln h}$. Explicitly, merging two ideal collections may modify the slope of the first part of Zipf's law.

According to the made assumptions, the number of ranks after rank $h$ is the same in $C_1 \cup C_2$ and in the two collections $C_1, C_2$. Therefore, the slope will not change in this section of the graph. However, for each distinct count, there will be a double number of elements with the same count, meaning that the low-rank end, shortly, the "broom" end of the corresponding Zipf's law graph will be twice as "fat".

Consequently, by merging two collections described by power laws with the same exponent, one obtains a noisy Zipf law, possibly with a fatter broom end and with slightly smaller slope at the starting tail.

When the exponents differ, the width of the "broom end" is affected, while the starting tail may have variable slope.

There is a variant of power law that we do not discuss here, namely, a generalized power law,

$$n(r) = \frac{A}{r^a} \cdot q(r) \qquad (12)$$

where $q(r)$ is a "fat queue" function.

IV. THE "BROOM"-END PROPERTIES OF ZIPF'S LAW

To the question: How many elements are in the group of rank $r$? Equ. (10) provides the answer.

Next, consider the question: How many ranks $r, r+1, \ldots, r+h$ have the same number of elements? (i.e., what is the number of ranks $r$ that actually have the same count?) When $r$ is small, no two successive ranks can have the same count, but for large values of $r$ we expect ranks with the same count; this implies that the distinction between ranks, for those elements, is conventional. One can see those groups of elements as having the same effective rank, where the effective rank is given by the count. If so, the elements corresponding to those effective ranks have to be of different type, else their counts sums and consequently, the joined elements will form a group that will have a different effective rank.

Consider the situations when two or several successive ranks produce the same count according to (10). That will produce in the graph of Zipf's law points at the same level $\ln(count)$. Denote by $\omega(r)$ the width of the broom-end at rank $r$. The width is given by the number of ranks with the same number of elements. If the number of elements is 1 (that is, the elements are hapax legomena, in linguistic applications), the condition comes to determine the maximum $r$ such that $A/r^a < 2$ (less than instead of less or equal because we will use the *floor* function in determining the number of elements of a given rank), moreover the maximum $h$ such that $1 \leq \frac{A}{(r+h)^a}$, that is,

$$1 \leq \frac{A}{(r+h)^a} < \frac{A}{r^a} < 2. \qquad (13)$$

The nontrivial conditions are

$$(r+h)^a \leq A, \ A < 2r^a \qquad (14)$$

or $(r+h)^a < 2r^a$, and so $r + h < 2^{1/a} r$, $h < \left(2^{\frac{1}{a}} - 1\right) r$. For $a = 1$, $h < r$. In general, for a count of $k$, the condition is

$$k \leq \frac{A}{(r+h)^a} < \frac{A}{r^a} < k+1. \qquad (15)$$

or, $k(r+h)^a < (k+1)r^a$, or $r + h < \left(\frac{k}{k+1}\right)^{1/a} \cdot r$, or $h < r\left(\left(\frac{k}{k+1}\right)^{\frac{1}{a}} - 1\right)$.

Taking the logarithm,

$$\ln h < \ln r + \ln\left(\left(\frac{k}{k+1}\right)^{\frac{1}{a}} - 1\right), \qquad (16)$$

Provided that at rang $r$ the count is k, that is $k = \frac{A}{r^a}$, $\ln k = \ln A - a \ln r$. One obtains a relation for $h$ as a function of the count $k$ or of the rank $r$, whichever is known,

$$\ln h < \frac{\ln A - \ln k}{a} + \ln\left(\left(\frac{k}{k+1}\right)^{\frac{1}{a}} - 1\right). \qquad (17)$$

Next, we are interested in a law approximating the width of the "broom"-tail for large values of the rank (toward the end of the "broom".

Let $M$ and $r$ be natural numbers and $a \geq 1, a \in \mathbf{R}$. How many natural numbers are between $M/r^a$ and $M/(r+1)^a$, assuming $M > r^a$? Let $a \in \mathbf{N}$, $n_1 = \frac{M}{r^a} - mod\left(\frac{M}{r^a}\right) \in \mathbf{N}$ and $n_2 = \frac{M}{(r+1)^a} - mod\left(\frac{M}{(r+1)^a}\right) \in \mathbf{N}$. Then, the number looked for is $n_1 - n_2$.

The number of integers between the successive ranks $n$ and $n+1$ for a power low with exponent -1 is then $v(n; M; a = 1) \leq \frac{M}{n \cdot (n+1)} \to M/n^2$. So, the maximal rank is $r_M \leq \frac{M}{n^2}$, because $r_M \geq 1$. From $\frac{M}{n^2} \geq 1$ one obtains $n^2 \leq M$, so, $r_M \leq \sqrt{M}$.

For the exponent $a = (-)2$, $\frac{1}{n^2} - \frac{1}{(n+1)^2} = \frac{2n+1}{n^2 \cdot (n+1)^2}$. The condition is $1 \leq \frac{2n+1}{n^2 \cdot (n+1)^2} < 2$, or $n^2 \cdot (n+1)^2 < 2n+1 \leq 2 \cdot n^2 \cdot (n+1)^2$.

The number of integers between the successive ranks $n$ and $n+1$ for the power law with $a = 2$ is $v(n; M; a = 2) \leq \frac{(2n+1)M}{n^2 \cdot (n+1)^2} \to \frac{2M}{n \cdot (n+1)^2} \sim \frac{2M}{n^3}$. So, the maximal rank is $r_M \leq \frac{2M}{n^3}$, because $r_M \geq 1$. One obtains $n^3 \leq M$, so, $r_M \leq \sqrt[3]{M}$. At the limit, $r_M = \sqrt[3]{M}$ and thus the number of hapaxes is $\frac{2M}{n^3}$

We conjecture that for whatever $a > 0$, there are two constants $\lambda, \Lambda$ such that the function $v(n; M; a) \xrightarrow[\substack{n \to \infty \\ M \to \infty}]{} \lambda n^\Lambda$.

**An approximation formula**

Let $M, r, s, q \in \mathbf{N}$, $a \in \mathbf{R}$, $s \leq \frac{M}{r^a} \leq s + 1$, $M = \lfloor sr^a \rfloor + q + \eta$, $0 \leq \eta < 1$. For example, $M = 16, r = 3, a = 1/2$; then $s = 9 < \frac{16}{\sqrt{3}} = 9.2376\ldots < 10$, and $q = 0$, $\eta = 0.2376\ldots$ Then,

$$\frac{M}{r^a} - \frac{M}{(r+1)^a} = M \frac{(r+1)^a - r^a}{r^a(r+1)^a} \qquad (18)$$

$$\frac{M}{(r+1)^a} + c = \frac{M}{r^a}, \ c = M \frac{(r+1)^a - r^a}{r^a(r+1)^a} \qquad (19)$$

$$\ln c = \ln M + \ln((r+1)^a - r^a) - \ln r^a - \ln(r+1)^a \qquad (20)$$

For large $r$, $\ln r \approx \ln(r+1)$. Then,

$$\ln c \approx \ln M + a \ln r + \ln\left(\frac{(r+1)^a}{r^a} - 1\right) - a \ln r - a \ln(r+1)$$

$$\ln c \approx \ln\left(M\left(\frac{(r+1)^a}{r^a} - 1\right)\frac{1}{(r+1)^a}\right) \quad (21)$$

For large $r$ and for not too small $a$, for example $a > 0.5$, $(r+1)^a - r^a \approx a(r+1)^{a-1}$. Then,

$$\ln c \approx \ln\left(M\left(\frac{a}{r}\right)\frac{1}{(r+1)^a}\right) \approx \ln\left(\frac{Ma}{(r+1)^a}\right) \quad (22)$$

Therefore

$$\nu(r; M; a) = \lfloor c \rfloor \approx \left\lfloor \frac{Ma}{(r+1)^{a+1}} \right\rfloor \quad (23)$$

and the number of integers between $\frac{M}{r^a}$ and $\frac{M}{(r+1)^a}$ is about $\left\lfloor \frac{Ma}{(r+1)^{a+1}} \right\rfloor$, with the approximation variant $\left\lfloor \frac{Ma}{r^{a+1}} \right\rfloor$. This proves a bound for the conjecture above.

As a matter of example, the error in the above computation is shown in Fig. 3.

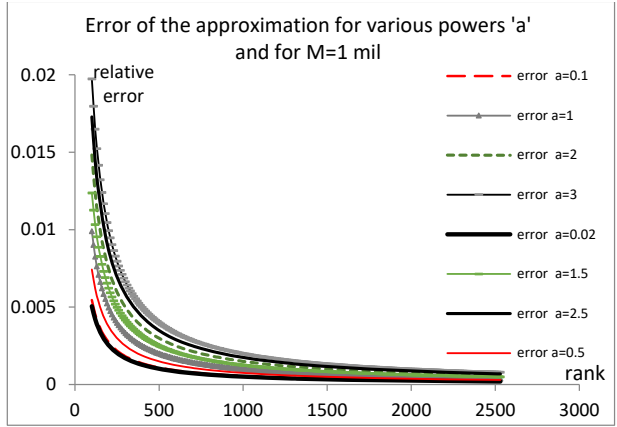

Fig. 3. Error in approximating the width of the broom end, according to equ. (23).

## V. RANK-DEPENDENT EXPONENT

We expose the case of rank-dependent exponent in the form $1/r^{(a+b/r)}$. This might be a model for populations where Zipf's law seems to be composed of two linear segments. Many texts and cases from economy exhibit such a behavior. Examples of graphs for various $b$ constants are shown in Fig. 4; they resemble the empirically determined "Zipf's law" for various processes, as found in the literature, e.g., Fig. 1 and 3 in (Williams et al.), Fig. 4 in (Neuwman 2006), Fig 3 in (Teodorescu M.H.M. 2017).

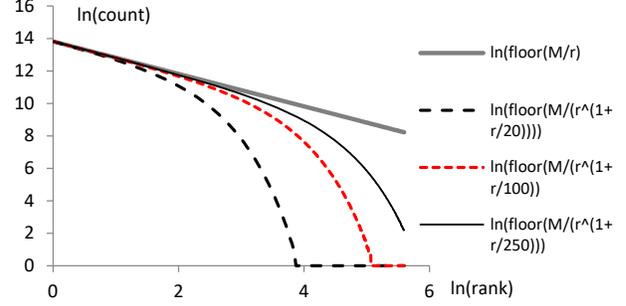

## VI. NOISE IN POWER LAW

First, notice that there is a natural "noise" added to the basic power law because we need to consider only integer counts; this "rounding" noise is $\left\lfloor \frac{N}{r_M^a} \right\rfloor - \frac{N}{r_M^a}$, which depends on the total number of words in the corpus and the maximal rank (which, in turn, is given by $\sum_{r=1}^{r_M} \frac{A}{r^a} = N$). Similar versions are obtained when applying ceiling or floor rounding functions. We are not concerned with this "noise".

It makes no sense to add noise to the ranks, because ranks are integers and therefore noise should be integer too, but then, adding noise would just change the range and correspondingly the count $n(r)$. One can easily add integer noise $\eta(r)$ to the counts as long as that does not change the range, that is,

$$n(r-1) + \eta(r-1) \geq n(r) + \eta(r) \geq n(r+1) + \eta(r+1) \quad (24)$$

These conditions combined with the power law lead to

$$\frac{A}{r^{a-1}} + \eta(r-1) \geq \frac{A}{r^a} + \eta(r) \geq A/r^{a-1} + \eta(r+1) \quad (25)$$

Or $\frac{A(r-1)}{r^{a-1}} \geq \eta(r) - \eta(r-1)$ and $\frac{A(r-1)}{r^a} \geq \eta(r+1) - \eta(r)$. As the noise may be supposed to be either positive or negative, a convenient, simplified condition is

$$\frac{1}{2} \cdot \frac{A(r-1)}{r^a} \geq \eta(r). \quad (26)$$

Essentially, the noise with the above conditions would not change the rank, but the count $n(r)$ can be anyway between $n(r-1)$ and $n(r+1)$.

Next, consider a more intuitive form of noise for the power law, the noise in exponent. The noise $\varepsilon(r)$ should obey some restrictions, namely $\frac{1}{r^{a+\varepsilon}} > \frac{1}{(r+1)^a}$, for the rank is preserved. For $r > 1$, this leads to $a \ln(r+1) > (a+\varepsilon) \ln r$, or $\frac{a+\varepsilon}{a} < \frac{\ln(r+1)}{\ln r}$, leading to

$$\varepsilon(r) = \varepsilon_0 \cdot a \cdot \left(\frac{\ln(r+1)}{\ln r} - 1\right) \quad (27)$$

where $\varepsilon_0$ is a random variable with values less than 1; a simple case is uniform noise $\varepsilon_0$. (Notice that $\frac{\ln(r+1)}{\ln r} - 1 =$

$\frac{\ln(r+1)-\ln r}{\ln r} \xrightarrow[r\to\infty]{} 0$). For example, $\varepsilon_0$ may be a uniform noise with the maximal absolute value less than 1.)

Other functions $g(r;a)$ may be chosen for noise shaping by multiplication, provided that they satisfy the condition $g(r;a) \leq a \cdot \left(\frac{\ln(r+1)}{\ln r} - 1\right)$ for all the ranks $r < r_M$.

In the context of computational linguistics, the noise amount in Zipf's law for a text or oeuvre may be indicative for the author's style.

## VII. A NOTE ON HAPAXES AND HONORÉ AND SICHEL INDEXES

Assuming that by adding text to another the probabilities of the words to appear do not change, any text addition will increase the numbers of distinct words without changing their rank (essentially). However, the words that where hapax may become duo and new hapaxes occur – namely, the words that are hapax in one but do not occur in the other text merged remain hapax in the merged text. Yet, NO new word becomes hapax, which was not already hapax in the merged texts. The probability that the same hapax occurs in both texts (before merging) is very low, because already the probability of a word being hapax in any of the texts is very low. What is the probability of hapax legomena, for a given $N, a$? What a significantly different number of hapax legomena tells us about the author, the corpus etc.?

Brooks (Brooks 2009), an author from Federal Bureau of Investigation, US, concluded her investigation about hapaxes as "The authorship attribution techniques of type/ token ratio, richness score, and hapax percentage were unsuccessful in distinguishing the performance of the SPs (suspect productions) from the rest of the corpus." As we proved, that is normal as several factors beyond authorship or style, specifically, number properties, decide what hapaxes are and how many they are.

Understanding how hapaxes occur in large texts is important for "the concept of idiolect and uniqueness of utterance" and in the analysis of authorship and plagiarism in courts and in forensic linguistics (Coulthard, 2004). The role of "the proportion of shared vocabulary" as a strong indicator of plagiarism may work well for short texts, as analyzed by has to be (Coulthard, 2004), but have limits when the text is large and the evidence is based on words rarely occurring (hapaxes, dislegomena, or similar), as the occurrences may be explained by probabilities and number theory. Stylometric indices, such as Honoré's index H = 100 × log N /(1 – (V1/V)), where "N is the number of word tokens in a target text, V1 is the number of hapaxes (words occurring only once) and V is the number of word types." (Baayen et al., 1996), (Ishihara), Brunettte index (Chandrika and Kallimani), and Sichel's measure (index), which takes into account the dislegomena, are also affected by the numerical artifacts discussed in the previous sections.

For large texts of two authors exhibiting a consistent power law distribution, with different exponents, the fact that there is a difference of the number of hapaxes and dislegomena in their texts is not an indicator of authorship or style, because the power law predicts that such a difference should exist. (We have fallen in this trap in (Teodorescu & Bolea, 2019) and in others). Only differences in the vocabulary (that is, of different hapaxes and dislegomena) may show a style difference. Of course, different inconsistencies in Zipf's law for their texts (when they have the same length), or different exponents of the power law may be indicative of their different style and of different authorship. instead of Honoré's index, which is largely determined by (23), we suggest that an index that is significant could be $I_{hp} = \frac{V_{hp} - \nu(r_M; M; a)}{\nu(r_M; M; a)}$, where $V_{hp}$ is the number of hapaxes in the text. This index requires the computation of $a$ and $r_M$; the index cannot be applied if the text does not follow a proper Zipf's law. Especially for small text, exact computation of $\nu(r_M; M; a)$ is preferable, instead of using (23). The proposed index will be negative for a meager vocabulary and positive for a rich one. A similar change is suggested for Sichel's index. Notice that the suggested indices measure departures from the theoretical Zipf's law. By extension, the total noise (sum of departures for all ranks) index is a generalization of the ones for hapaxes and dislegomena.

## VIII. DISCUSSION, FURTHER QUESTIONS, AND CONCLUSIONS

Numerous questions and potential developments in the same line remain unaddressed. Consider a function $h(n): N \to R$ that strictly decreases to zero. Define a function similar to zeta function by $\xi(h) = \sum_{k=1}^{\infty} h^n(k)$; in particular, $h(n) = \frac{1}{n^a}, a > 0$ (or $a \in C$). The problem is to determine the function $\nu(n; M; h)$ of the integers between two successive values of $n$, this is a problem similar to that discussed for zeta function.

Language models may have a predictable behavior of Zipf's law, for texts from various domains and for various text lengths. It is unclear if an AI machine, such as ChatGPT or similar, can be identified as an 'author' by the associated Zipf's law, or by derived indices (hapaxes, Sichel, Honoré etc.). However, in an adversarial setting, these models can be trained or adapted, e.g., taught by rules, to modify during time its specific power law, for example to deviate from a basic power law using noise or to randomly hop through a set of exponents.

In case of a process with probabilities given by the power law, $p(r) \sim 1/r^a$, the estimated number of elements of rank $r$ is $M/r^a$, $M$ a constant, leading (in average) to the cases already discussed.

For models involving other distributions that are similar to the power law, for example the lognormal distribution, the presented analysis should be developed in the future.

Further examples of the applications of Zipf-like laws are discussed in (Teodorescu M., 2023). A specific topic of discussion is presented in Annex 1.

Among others, the discussion in this article shows that the method always used for determining the approximative power law from the empirical data, where the method is based on minimizing the square error, is inappropriate; the subject will be dealt with in another article.

### ACKNOWLEDGMENT

Special thanks: Prof. Mike HM Teodorescu suggested that the 'broom'-end of the graphs of Zipf's law needs a deeper study, including the investigation of its 'noise' mechanisms, assisted with the derivation of most of the formulas, has proposed the use of bootstrapping in the analysis of style differences for different corpora, proposed the use of Jaccard, cosine and distances for comparing the vocabularies of the hapaxes and dislegomena and for determining how much noisy power law series can differ from the ideal one in terms of 'broom' end, and proposed several ideas found throughout the paper and particularly in the first and last section.

**Annex 1**

We are interested in the properties of the function $\zeta_n(a, M) = M \sum_{r=1}^{n} \frac{1}{r^a}$, $M$ a large positive number, $a > 0$. Specifically, we are interested in the values $n$ where, for some $m < M$ natural number, $m < \zeta_n(a, M) \leq m + 1$. Precisely, we denote

$$\varphi(m; M) = |\{n | m < \zeta_n(a, M) \leq m + 1\}|$$

and are interested in determining this function, where the function value for some $m$ represents the number of partial sums $\zeta_n(a, M) = M \sum_{r=1}^{n} \frac{1}{r^a}$ that have values between $m$ and $m + 1$. The function $\varphi$ represents the width of the tail of some set of $M$ objects with power law distribution. Similar functions can be defined based on the series of other probability density or cumulative functions.

More generally, for a function $\eta(n, a, M)$ modeling noise of the $\zeta_n(a, M)$ function, we may be interested in

$$\varphi_\eta(m; M) = |\{n | m < \zeta_n(a, M) + \eta(n, a, M) \leq m + 1\}|$$

The functions $\varphi(m; M)$ and $\varphi_\eta(m; M)$ have the property that

$$\sum_{m=0}^{M} \varphi(m; M) = \sum_{m=0}^{M} \varphi_\eta(m; M) = M.$$